%% file: main.tex
\RequirePackage{fix-cm}
\documentclass{svjour3}                     
\smartqed  
\usepackage{graphicx}
\usepackage{amssymb}
\usepackage{amsmath}
\usepackage[numbers]{natbib}

\newcommand{\subparagraph}{}
\usepackage{titlesec}
\let\subparagraph\relax
\journalname{The Journal of Supercomputing}

\usepackage{listings}
\lstnewenvironment{Code}[1][]{
\lstset {
     float=htp,
     frame=single,
     framextopmargin=2pt,
     framexbottommargin=2pt,
     basicstyle=\scriptsize\ttfamily,
     breaklines=true,
     language=c++,
     morekeywords={*,pragma, rolex},
     #1
  }
} {}

\lstnewenvironment{CodeExample}[1][]{
\lstset{
     float=h,
     frame=single,
     basicstyle=\scriptsize\ttfamily,
     breaklines=true,
     commentstyle=\textit,
     framextopmargin=2pt,
     framexbottommargin=2pt,
     morekeywords={*,rolex, resilience, robust, share, private, compare},
     #1
  }
} {}

\title{Rolex: Resilience-Oriented Language Extensions for Extreme-Scale Systems 
\thanks{The authors would like to acknowledge the support for this work provided through Scientific Discovery through Advanced Computing (SciDAC) program funded by U.S. Department of Energy, Office of Science, Advanced Scientific Computing Research under award number DE-SC0006844. Partial support for this work was also provided by the US Army Research Office (Award W911NF-13-1-0219)}}

\author{Saurabh Hukerikar, Robert F. Lucas}

\institute{Information Sciences Institute \at
	   University of Southern California \\
	   4676 Admiralty Way Suite 1001, \\
	   Marina del Rey, CA 90292 USA \\
           \email{saurabh, rflucas@isi.edu}           
}

\date{Received: date / Accepted: date}

\begin{document}
\maketitle

\begin{abstract}
\input{00_Abstract}

\end{abstract}

\input{01_Introduction}
\input{02_ProgrammerKnowledge.tex}

\input{03_LangExtDesign.tex}

\input{04_Syntax_Semantics.tex}

\input{05_Compiler_Runtime.tex}

\input{06_Experimental_Eval.tex}

\input{07_Related_Work.tex}

\input{08_Conclusion.tex}


\bibliographystyle{spbasic}       
\bibliography{main}               

\newpage
\appendix 
\input{A_Appendix_Grammar.tex}

\end{document}

%% file: 00_Abstract.tex
Future exascale high-performance computing (HPC) systems will be constructed from VLSI devices that will be less reliable than those used today, and faults will become the norm, not the exception. This will pose significant problems for system designers and programmers, who for half-a-century have enjoyed an execution model that assumed correct behavior by the underlying computing system. The mean time to failure (MTTF) of the system scales inversely to the number of components in the system and therefore faults and resultant system level failures will increase, as systems scale in terms of the number of processor cores and memory modules used. However every error detected need not cause catastrophic failure. Many HPC applications are inherently fault resilient. Yet it is the application programmers who have this knowledge but lack mechanisms to convey it to the system.

In this paper, we present new Resilience Oriented Language Extensions (Rolex) which facilitate the incorporation of fault resilience as an intrinsic property of the application code. We describe the syntax and semantics of the language extensions as well as the implementation of the supporting compiler infrastructure and runtime system. Our experiments show that an approach that leverages the programmer's insight to reason about the context and significance of faults to the application outcome significantly improves the probability that an application runs to a successful conclusion.

%% file: 01_Introduction.tex
\section{Introduction}
\label{section:Introduction}

By the end of this decade, exascale high-performance computing (HPC) systems promise to accelerate the pace of scientific discovery in a broad range of disciplines including climate and environmental modeling, chemistry and materials, high energy and nuclear physics, nanotechnology, astrophysics, and biology. These systems will enable the solution of vastly more accurate predictive models and the analysis of massive data sets \cite{ASASC:2010}. 

Among the difficult challenges in designing and operating future exascale-class systems, guaranteeing reliability of operation in the presence of increasingly frequent faults and errors will be critical. Various studies \cite{Dongarra:2011:IES}\cite{DARPA_ExascaleTechStudyReport:2008} have suggested that the path to higher capability machines will require an exponential increase in the number of CPU cores and memory modules in order to drive performance. For an exascale-class supercomputer, its sheer scale is a challenge to the system's ability to tolerate faults and maintain service. Furthermore, the reliability of individual components is projected to decrease as Moore's law enables shrinking transistor geometries \cite{DARPA_ExascaleResilienceStudyReport:2009}.

In today's HPC systems, we enjoy a model of execution in which the application presumes correct behavior by the underlying fabric of hardware and system software, i.e., the execution environment. Some errors are masked by hardware-based mechanisms, and the error events that cannot be handled by the system layers usually result in fatal crash. This is usually catastrophic for the application processes running on the system. Therefore most HPC systems deal with anomalous events only when they result in catastrophic failure through a process of checkpoint and rollback recovery (C/R). However, for the projected fault rates in future exascale-class HPC systems relying solely on such mechanisms will lead to frequent application failures or incorrect results. 
Many of the scientific applications that run on these systems contain features that allow the effect of certain faults and errors to be tolerated or mitigated at the application level through algorithmic methods. Various algorithm-based fault tolerance (ABFT) solutions \cite{Huang:1984} \cite{Bosilca:2008} support application-level error detection and correction. Therefore, not all faults and errors need to result in a catastrophic crash. Programmers of scientific applications, through their domain expertise and familiarity with the application codes, gained through code optimization efforts, are usually well-positioned to understand such application-level fault-resilience features. However, they lack convenient mechanisms to express such knowledge to the system. We believe that with modest extensions to existing programming model the application-level knowledge may be leveraged by the execution environment to enable HPC applications to continue running towards successful completion despite the presence of certain faults and errors in the system. In this paper we investigate whether simple language-level extensions in concert with a compiler infrastructure and a runtime inference framework can enhance the ability of HPC applications to manage the effects of faults and errors in their state. We propose Rolex, a set of \textbf{R}esilience-\textbf{O}riented \textbf{L}anguage \textbf{E}xtensions that capture HPC programmers' knowledge of the fault-tolerance features of the program code and their expectations of application outcomes. By making resilience essential to the programming model, the execution environment can use this application-level knowledge to reason about the significance of the errors to the correctness of the application's outcome. We define the syntax of the resilience-oriented language extensions, describe their fault-resilience semantics, and their integration with a compiler infrastructure and runtime inference system. We also describe our experience of applying Rolex to several common HPC application codes and evaluate the application resilience using accelerated fault injection experiments.

The remainder of this paper is organized as follows: Section \ref{sec:ProgrammerKnowledge} explains the basis of our approach on how capturing programmer knowledge through simple language extensions may be used to manage the application’s fault resilience. Section \ref{sec:LangExtDesign} describes the design goals and philosophies behind the Rolex extensions and Section \ref{sec:Syntax_Semantics} presents their syntax and semantics and several motivating examples which demonstrate the viability of applying these language extensions in the context of real HPC applications. Section \ref{sec:Compiler_Runtime} elaborates the role of the compiler and runtime inference engine. Section \ref{sec:Experimental_Eval} presents the evaluation results for fault injection experiments and also studies the impact on application performance. Section \ref{sec:Related_Work} surveys related programming model-based resilience approaches.


%% file: 02_ProgrammerKnowledge.tex
\section{Leveraging Programmer Knowledge for Fault Resilience}
\label{sec:ProgrammerKnowledge}

The HPC workload consists of scientific computations, many of which are naturally tolerant to data errors. Their algorithmic behavior might simply filter the occasional incorrect value, as is the case with many numerical iterative algorithms, or they might rely on pseudorandom processes, as is the case with Monte-Carlo techniques. Several applications that use numerical analysis methods can tolerate limited loss in floating point precision. In certain applications, the impact of errors in the data or computation can even be trivially healed through simple algorithmic methods. For example, parity and checksums can be applied to specific data structures or procedure executions to detect the presence of data corruptions within the application's address space. However, part of the variable state, especially that which affects program control flow and pointer arithmetic, is very sensitive to errors. Therefore, for certain parts of the program state, the notion of correctness may be defined within the bounds of certain rounding error, while for others it may require precise bit reproducible correctness \cite{DoE:ResilienceReport}. 

\begin{figure} [t]
\centering
\includegraphics[width=\linewidth]{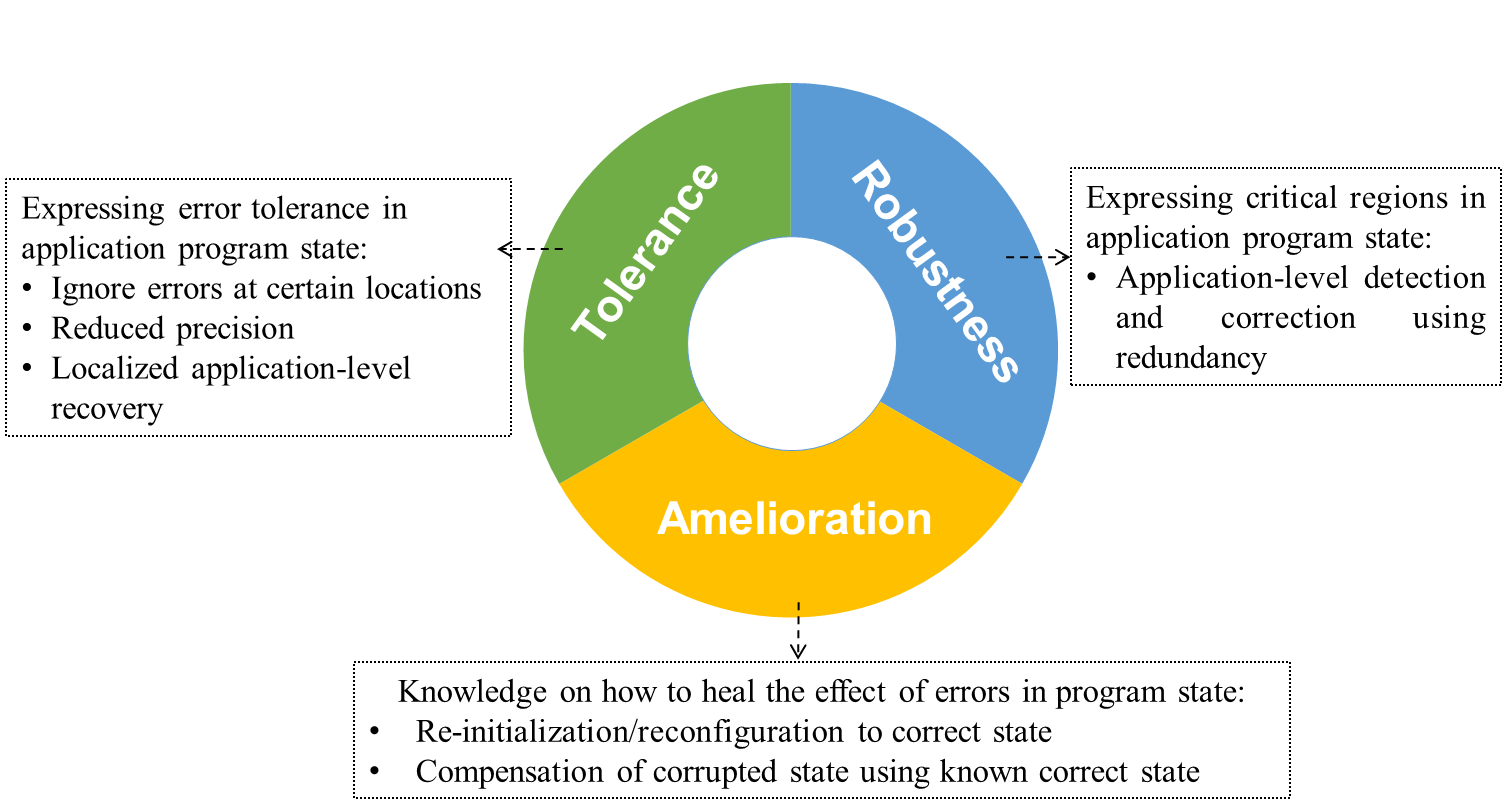}
\caption{Themes of programmer knowledge to enhance application resilience}
\label{Fig:ResilienceKnowledge}
\end{figure}

HPC application programmers are well-positioned to understand the application's fault-tolerance features because they tend to be experts in their respective scientific domains and due to their familiarity with the program code structure. We believe that given appropriate interfaces to express their fault tolerance knowledge, programmers can contribute to enhancing the execution environment's management of the application resilience. Through programming model features we may be able to support fault-tolerance capabilities, namely error detection, containment and recovery at the application level. Such programming model-based mechanisms provide a fine-grained model of reliability in which individual data variables and program statements may be tuned for relaxed or strict reliability and seek to prevent application failure for every possible error instance in the system.

Broadly, the knowledge that programmers can express falls into three major themes. These are illustrated in Figure \ref{Fig:ResilienceKnowledge} along with plausible solutions and described below:
\begin{itemize}
\item \textbf{Tolerance:} A programmer may choose to tolerate limited loss in floating precision for certain program values, or allow occasional perturbations of certain data values. The programmer may also be aware of regions of computation that employ iterative refinement, such that errors which cause anomalous intermediate results may be absobred without affecting the correctness of the final outcome.
\item \textbf{Robustness:} Certain data structures and computation, notably those related to the program control flow and pointer arithmetic need bit-level correctness. The programmer may identify application-level constructs that require stronger checks. The error detection and correction may be accomplished by maintaining redundant copies and using masking mechanisms to guarantee deterministic program behavior.
\item \textbf{Amelioration:} A variety of algorithmic techniques exist that not only detect but also heal the effect of errors in data structures. Such techniques maintain redundant information, such as checksums, to recover erroneous values. They may also use value re-initialization to repair variable state. Certain computations even allow compensating erroneous values by interpolating neighboring values. The programmer may be able to provide the appropriate methods to ameliorate program state.
\end{itemize}

Programming model extensions designed to enable the execution environment to capture application-level features on each of these themes of knowledge supports a fault-aware execution environment that can provide error resilient operation for HPC application processes without compromising the application performance, or the productivity of programmers.

%% file: 03_LangExtDesign.tex
\section{Design of the Resilience Oriented Language Extensions}
\label{sec:LangExtDesign}

\subsection{Goals for Resilience-Oriented Language Extensions}
In designing the language extensions, we sought to capture each of the flavors of knowledge described in Section \ref{sec:ProgrammerKnowledge}, and in the process, also enable each of the aspects of fault management namely detection, containment and recovery. Broadly, our goals for the resilience-aware programming model extensions are:

\begin{enumerate}
\item It is our goal to retain the familiarity of current programming paradigms. We aim to adopt a simple syntax that permits embedding resilience capabilities within existing programming language features.

\item We seek to minimize the time and effort required by programmers to learn and adopt the language extensions; therefore, these resilience-oriented language extensions must provide a concise and elegant syntax and include a small set of new language keywords for expressing the resilience features.

\item We also seek a fair division of work between the language extensions and the compiler and runtime framework, such that the programmer does not need to be exposed to the complexity of the HPC execution environment, yet is provided with sufficient abstractions to be able to concisely convey fault management knowledge related to application-level constructs.

\item Recognizing that HPC programmers are very reluctant to trade off their performance, which is usually achieved by investing much time and effort in hand-tuning the code, we seek to ensure that the resilience-oriented language extensions and compiler transformations do not drastically affect the code structure.

\item As HPC systems become increasingly heterogeneous and topologically complex in pursuit of higher performance, they need to harness a variety of novel parallel programming frameworks. Yet the applications seek to retain the well-understood foundation of the Message Passing Interface (MPI) as well as certain well-tuned productivity libraries such as BLAS and LAPACK written in C and FORTRAN. It is also our goal to ensure that resilience-oriented language extensions integrate seamlessly with these language features and library frameworks.
\end{enumerate}

\subsection{Description of Syntactic Structure of Rolex}
Based on these objectives we have designed programming language extensions that include a collection of features that extends the base language as well as compiler directives and runtime library routines that enable the execution environment to manage the application's error resilience. Rolex is designed to affect the following aspects of the program state \cite{Langou:2007}: (i) the \textit{computational environment}, which includes the data needed to perform the computation, i.e., the program code, environment variables etc.; (ii) the \textit{static data}, which represents the data that is computed once in the initialization phase of the application and is unchanged thereafter; (iii) the \textit{dynamic data}, which includes all the data whose value may change during the computation. Rolex extends the C, C++ language with constructs that provide application-level error detection, error containment and recovery strategies for each of these aspects of the program state. These extensions fully comply with the syntactic structure of the base language grammar and complement the existing language features. Some Rolex constructs serve as directives for the compiler to automatically generate code that supports fault resilience, whereas the Rolex routines support application-level resilience through the runtime environment.

\subsubsection{Type Qualifiers}
Rolex extends the declaration ability of C/C++ to allow type qualifiers that enable attaching a specific resilience attribute to functions, data variables and other objects. The programmer specifies, through explicit association, an error detection and/or tolerance feature for specific identifiers in the program code. The syntatic structure for the use of resilience type qualifiers is:

\begin{Code} 
 <rolex-error-management-qualifier> variable-declaration;
\end{Code}

The formal rules that extend the C/C++ grammar to include the resilience-oriented type qualifiers are described in Appendix A in Listing \ref{lst:TypeQualifierRules}. Through these qualifiers, the programmer explicitly specifies how the program variables are managed, when the associated object value is deemed to be in erroneous state. The error detection and correction capabilities are handled through bit manipulation on the low-level representation of the objects.

\subsubsection{Directives}
\label{sec:LangExtDesign_subsec:Directives}
Rolex directives enable the application programmer to impose rules for fault-tolerant execution of a region of the program code. In C/C++, {\tt \#pragma} directives specify program behavior. The syntactic structure of an executable Rolex directive and the code region is: 
\begin{Code} 
#pragma rolex <error-management-directive> [clause[[,] clause] ... ] new-line
{
    /* binding region: structured blk*/
}
\end{Code}

The \textit{binding region} determines the scope of the execution context that is equipped with resilience capabilities. The bound region is a structured block, which is defined as an C/C++ executable statement, which may be a compound statement but has a single point of entry at the top and single point of exit at the bottom. The compound statement is enclosed within a pair of \texttt{\{} and \texttt{\}}. The point of entry cannot be the target of a branch and the point of exit cannot be a branch out. No branch is allowed from within the structured block, except for program exit.  Instances of the structured block may be compound statements including iteration statements, selection statements, or try blocks.

We also provide declarative directives that may be associated with function declarations and definitions:
\begin{Code} 
#pragma rolex declare <error-management-directive> [clause[[,] clause] ...] new-line
  /* C/C++ function definition or declaration */
\end{Code}

These directives are not associated with the immediate execution of the application code but enable the compiler to create multiple versions of the specified C/C++ function, at least one of which includes resilience capabilities. The Listing \ref{lst:DirectiveRules} in Appendix A shows the grammar rules for the extensions based on the resilience-oriented directives.

\subsubsection{Runtime Library Routines} 
Certain aspects of the resiliency of the execution environment can be controlled through runtime library routines. Also, some of the existing standard library calls may be extended to provide resilience capabilities. For example, the memory management library calls are equipped with error detection, correction and recovery capabilities on the allocated memory blocks. The routine identifier is suffixed with the fault management capability: 
\begin{Code}
return_type var = rolex_libraryfunc_capabilitity ( 'arguments' );
\end{Code}

These routines are external C functions whose identifiers are prefixed with a {\tt rolex} keyword. 

\subsubsection{Rolex Keywords}
We introduce a set of keywords that are distinct from the existing set of C/C++ reserved keywords in order to support resilience semantics on the C/C++ constructs. The Rolex directives and routines are identified by the {\tt rolex} keyword. Additionally, the keywords {\tt tolerant}, {\tt robust}, {\tt heal} are used as qualifiers in type declarations. The keywords {\tt recover-rollback} and {\tt recover-rollforward} are used to associate a recovery behavior to a structured code block following a directive while the keyword {\tt robust} is used to specify redundancy in state or computation. Additionally, there are clauses that support management of variable state and permit specification of the strength of redundancy in the context of Rolex constructs.

%% file: 04_Syntax_Semantics.tex
\section{Rolex: Syntax and Semantics}
\label{sec:Syntax_Semantics}

This section provides more complete lexical syntax (how these extensions may be embedded in real programs) based on the syntactic structure from the previous section. The extensions support each of the previously described themes of knowledge, i.e., tolerance, robustness and amelioration. We also explain the semantics (what each extension means), how Rolex features affect program structure and their relationship to the runtime system. We also provide motivating examples that demonstrate how each Rolex feature enables fault resilience in real scientific application codes.

\input{04_a_Syntax_Semantics_TOLERANT}

\input{04_b_Syntax_Semantics_ROBUST}

\input{04_c_Syntax_Semantics_AMELIORATION}

%% file: 04_a_Syntax_Semantics_TOLERANT.tex
\subsection{Tolerance-based Extensions}
The \textbf{tolerance} language extensions are used to specify data variables or code block executions that support \textit{error elision}, i.e., ignore the presence of a corruption in program state and continue execution with the confidence that the algorithm can absorb the error or mask it through localized recovery. The extensions also enable applications to continue execution with imprecise but not unreasonable state through value coercion \cite{Hukerikar:FTXS:2012}. The extensions assume that error detection is provided by the hardware or system software, and that the error notification is communicated to the runtime system via an interrupt mechanism. 

For errors detected that happen to be mapped to locations that have been explicitly specified as tolerant using Rolex, the runtime system reacts to an error notification by allowing an application execution to continue despite the corruption in its state. For instances of errors that are mapped to locations on which tolerance is not specified, the runtime terminates the application execution, as is the standard behavior for unrecoverable errors.

\subsubsection{Type Qualifiers}
\paragraph{Syntax} \hspace{0pt} \\
The \texttt{tolerant} type qualifier can be applied to primitive as well as compound data structures. These qualifiers can be applied to declaration of global variables and local automatic variables and may include static and dynamic program state. 
The syntax for the type qualifiers variable declarations is:
\begin{CodeExample}[label={lst:Example1},frame=single]
 tolerant(PRECISION=...) float low_precision_32;

 tolerant(PRECISION=...) double low_precision_64;
 
 tolerant unsigned int rgb[X_RES][Y_RES];

 tolerant (MAXIMUS = 1023) unsigned int counter;
\end{CodeExample}

For floating point variables, the qualifier contains an additional specifier for precision. For integer values, the qualifier contains an additional specifier for maximum value. 

\begin{figure} [ht]
\centering
\includegraphics[height=10mm,width=\linewidth]{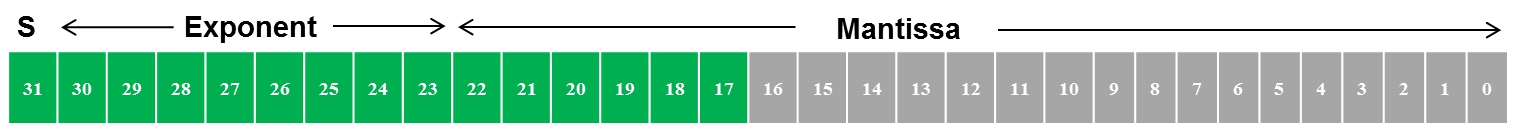}
\caption{IEEE 754 floating point representation}
\label{Fig:IEEE_754_FP_Representation}
\end{figure}

\begin{figure} [ht]
\centering
\includegraphics[height=7mm,width=\linewidth]{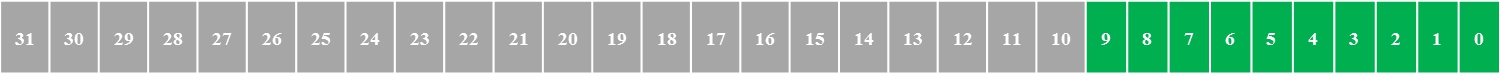}
\caption{Unsigned integer (32-bit) representation}
\label{Fig:Integer_Representation}
\end{figure}

\paragraph{Semantics} \hspace{0pt} \\
With these type qualifiers, error elision is achieved through coercion of the object value. For floating point objects, bit perturbation errors on the sign and exponent bits fundamentally alter the variable value, and the application is usually intolerant to such errors (shown in green in Figure \ref{Fig:IEEE_754_FP_Representation}). However, bit perturbations in the lower significand/mantissa bits may be ignored by the runtime and result in a truncation error in the value of the floating point variable (shown in grey in Figure \ref{Fig:IEEE_754_FP_Representation}). The {\tt PRECISION} construct specifies the minimum floating point precision that the programmer expects, i.e., it indicates the amount of precision loss the programmer is willing to tolerate. 
For an integer variable whose maximum value is known apriori, only the lower significant bits in the bit representation are intolerant; i.e., these bits cannot accept bit perturbations without altering the value of the variable (shown in green in Figure \ref{Fig:Integer_Representation}). The upper significant bits are unused and are meant to always remain '0' (for unsigned integers in the binary representation). When these bits are perturbed, the error may be masked by simply resetting these bits. This knowledge may be explictly conveyed through the {\tt MAXIMUS} construct in the type qualifier.   

The runtime responds to notifications that indicate the presence of an error which is mapped to a tolerant qualified data variable by manipulating the bit representation to coerce the data values into lower precision or mask the anomalous bits and allows the application execution to resume. These type qualifiers offer the program variables with error containment and limited recovery capabilities by masking perturbations and keeping their the values within permissible range of correctness.

\subsubsection{Directives} \label{subsec:Tolerance:Directives}
\paragraph{Syntax} \hspace{0pt} \\
The tolerance directives provide limited localized recovery capability from errors in the computation for the programmer-defined code regions. When the detected error maps to code sections, i.e., instruction memory of the application address space, or to the variables manipulated by the code region, the tolerance directive offers roll-back and roll-forward capabilities for the affected structured code block. The syntax of the tolerance roll-forward and roll-back directives is:
\begin{CodeExample}[label = {lst:Example3},frame=single]
#pragma rolex recover-rollback share ( variable_list ) private ( variable_list )
{    /* code block */    }

#pragma rolex recover-rollforward share ( variable_list ) private ( variable_list )
{    /* code block */    }
\end{CodeExample}

In order for the program state to remain consistent upon roll-forward/roll-back, the variable state must be the same as that during initial entry into the code block. Therefore, we provide optional \texttt{share} and \texttt{private} clauses that list the variables that need to be preserved and restored. 

The {\tt declare} directives instruct the compiler to generate versions of the associated functions with retry or ignore capabilities. The syntax (shown below) contains an optional {\tt fallback} clause to specify a default function return value. 
\begin{CodeExample}[label = {lst:Example4},frame=single]
#pragma rolex declare resilient ignore fallback() 
    /* function definition or declaration */
#pragma rolex declare resilient retry fallback()
    /* function definition or declaration */
\end{CodeExample}

\paragraph{Semantics} \hspace{0pt} \\
When the runtime is informed of the presence of an error that is mapped to the instruction memory of the \textit{tolerant} structured code block, or to one of the data structure variables specified in the data clauses, the structured block is re-entered (the execution is rolled back) or the remaining code block is skipped (the execution is rolled forward). The initiation of roll-forward or roll-back may cause the data variable state to become inconsistent. Therefore, prior to original entry into the structured code block, the variables specified in the \texttt{share} clause are saved. Upon roll-forward or roll-back recovery, this variable state is restored to the previously preserved values. The variables in the \texttt{private} clause are not restored and are treated much like local automatic variables declared inside a function. The declare directives allow the qualified execution to be retried, or it may be discarded with the function caller receiving a default fallback value. The tolerance directives offer error containment by limiting the scope of error to the computation contained in the block following the directive. Additionally, these directives also support compensation-based recovery of the application's variable state and localized recovery of erroneous computation through roll-forward/roll-back semantics.

\subsubsection{Runtime Library Routines}
\paragraph{Syntax} \hspace{0pt} \\
The Rolex tolerant routine extends the functionality provided by malloc(). It accepts an additional parameter of type \texttt{rolex\_precision} to specify the \texttt{MAXIMUS} and \texttt{PRECISION} for individual primitive types (when the routine is used to allocate arrays of primitive integer or floating point type). The API format is: 
\begin{CodeExample}[label = {lst:Example5},frame=single]
 float* intermediate_sol_array =  (float*) rolex_malloc_tolerant ( N * sizeof (float), NULL );

 float* molecule_position =  (float*) rolex_malloc_tolerant ( N * sizeof (float) , (rolex_precision) (6) ); /* PRECISION = 6 */

 unsigned int* true_color_pixel_buffer =
(unsigned int*) rolex_malloc_tolerant ( N * N * sizeof (unsigned int),  (rolex_precision) (16,777,216) ); /* MAXIMUS = 16,777,216 */
\end{CodeExample}

\paragraph{Semantics} \hspace{0pt} \\
Much like the standard library malloc, the {\tt rolex\_malloc\_tolerant()} allocates a block of memory whose address bounds are registered with the runtime system. Since such error-tolerant memory is explicitly requested, the runtime supports error elision, i.e., it ignores the notifications of any errors detected on this memory block and allows the application execution to resume.
For compound data structures composed of floating point or integer primitive types, the argument of {\tt rolex\_precision} type supports elision through value coercion, i.e., it allows the application to respond to error notifications by resuming the execution after ensuring the individual floating point or integer data values meet the precision or maximum values specified in the {\tt PRECISION} or {\tt MAXIMUS} constructs.

\subsubsection{Examples} \hspace{0pt}
Scientific modeling entails representation of continuous problems in terms of finite precision values which incurs some discretization error. Certain data structures in these applications may accept bit perturbations that result in round-off errors without affecting the validity of the simulation. Numerical analysis algorithms, such as the conjugate gradient method and the generalized minimal residual method (GMRES), progressively improve an initial approximate solution and terminate only when the solution is below a certain error norm. Direct methods such as Gaussian elimination and the QR factorization method terminate in a finite number of steps, but still yield an approximate solution. Limited loss in floating point precision in the intermediate solution state may be absorbed without impacting the correctness of the final solution.   

Molecular dynamics (MD) simulations can maintain the numerical stability with limited loss in floating point precision for various constant energy and constant temperature simulations. The deviations in the force calculations are often small enough that the particle trajectories are almost identical in terms of numerical stability as full precision calculations. In large scale simulations the loss in precision in lower significand floating point bits results in a negligible difference in the coordinates of the simulations over millions of time steps \cite{Zou:2012}. The Hartree-Fock method, used in computational chemistry codes, contains structures such as the Fock matrix, density matrix, matrix exponential, and orbital transformation matrix, which can tolerate bit perturbations in the lower significant mantissa bits in the mantissa of floating point representation \cite{vanDam:2013}. Such structures may be tolerant type qualified or allocated using \texttt{rolex\_malloc\_tolerant()}. Similarly, visualization applications allow arbitrary bit flips on integer type pixel values because the graphics rendering pipeline often accounts for incorrect pixel attributes.   

Algorithms that permit selective reliability may utilize directives to specify fault tolerant behavior for application phases. The FT-GMRES algorithm \cite{Hoemmen:2011} uses inner-outer iterations where the inner solver step preconditions the outer iteration. The inner solver step may be treated as an unreliable phase since it is allowed to return an incorrect solution without affecting the outer solver step. Similarly, neutron transport (NT) simulation codes use the Monte Carlo method and we may leverage its stochastic nature along with the fact that the simulation of every particle is independent. The code regions that create and simulate individual particles may be included in the structured block following tolerant directives, which allows the simulation to selectively discard the particles that experienced errors.

%% file: 04_b_Syntax_Semantics_ROBUST.tex
\subsection{Robustness-based Extensions}
The \textbf{robustness} language extensions are used to specify data variables or code blocks that are critical to the application correctness and as such could benefit from error detection and correction at the application-level. These include the application code sections (i.e., instruction memory), pointer variables, array index references as well as variables that affect control flow decisions. These aspects of the program state require bit-precise correctness in order to make a deterministic assertion on the correctness of the application outcome, even if it runs to completion in the presence of program state corruptions (but without raising any exceptions or abnormal behavior).
The robustness of these aspects of the program state may be guaranteed by the use of redundancy. This entails replicating part of the variable state, or specific portions of the program code execution, or at times both. The replicated part of the program state is compared to check for the presence of errors in the application's address space, or to filter errors through majority voting. Through these Rolex extensions the redundancy is selectively applied only on the sensitive data variables and computation whose correctness is critical to produce a correct application outcome.

\subsubsection{Type Qualifiers}
\paragraph{Syntax} \hspace{0pt} \\
The \texttt{robust} type qualifier may be applied to declarations of primitive as well as compound data structures. The syntax for the robust type qualifier, which includes a strength clause, is:
\begin{CodeExample}[label = {lst:Example6},frame=single]
 robust (CORRECT) int* csr_matrix[row_offsets];

 robust (DETECT) int* graph_edge_list[N];
\end{CodeExample}
\paragraph{Semantics} \hspace{0pt} \\
The type qualifier serves as a directive to the compiler, which performs source-to-source translation to duplicate or triplicate the variable declaration. For pointer variables, this amounts to creating aliases to the object being referenced. The compiler also duplicates/triplicates the statements in the program source that operate on the robust qualified variables as well as inserts statements that compare the redundant variable values and report any mismatch among the replicas to the runtime system. The qualifiers enable error detection and correction capabilities on the robust qualified objects and implicitly on their computation through statement-level DMR or TMR. 

\subsubsection{Directives}
\paragraph{Syntax} \hspace{0pt} \\
The robust directives provide application-level detection/correction for specific regions of computation, whose scope is defined by the structured code block following the directive. The declarative robust directives may be applied to functions. The syntax for the directives is: 
\begin{CodeExample}[label = {lst:Example7},frame=single]
#pragma rolex robust detect share ( variable_list ) private ( variable_list ) compare ( variable_list )
{   /* code block */  }

#pragma rolex robust correct share ( variable_list ) private ( variable_list ) compare ( variable_list )
{   /* code block */  }

#pragma rolex declare resilient robust (detect) fallback() 
    /* function definition or declaration */
\end{CodeExample}

The directives contain a \texttt{strength} clause, which specifies whether DMR or TMR is required for the structured block. The data management clauses \texttt{share} and \texttt{private} specify the data-sharing attributes for the variables listed in the respective clauses. The \texttt{compare} clause is used to specify the list of variables produced by the structured blocks that need to be compared/majority voted on to detect/correct an error in the computation. The fallback clause is used to return a default value to the function caller when the redundant execution of the function detects an error but is unable to conclusively vote on a correct value.

\paragraph{Semantics} \hspace{0pt} \\
When the compiler encounters the robust directive, it outlines the application code contained in the structured code block. It inserts statements that enable the redundant execution of the outlined code block by duplicating or triplicating the call to the outlined function and statements to compare the outputs of the structured block. The compiler also selectively replicates the variables in the data scoping clauses. Each redundant code block instance owns a separate replicated copy of \texttt{private} variables whereas a single copy of \texttt{share} scoped data is accessed by all redundant code block copies with the programmer responsible for synchronized access. The robust directives provide error containment by limiting scope to computation contained in the structured block in addition to the detection and correction capabilities.

\subsubsection{Runtime Library Routines}
\paragraph{Syntax} \hspace{0pt} \\
The robust version of the memory allocation routine supports redundancy-based error detection and/or correction for the dynamically allocated memory on the heap section of the application address space.  The routine prototypes are:
\begin{CodeExample}[label = {lst:Example9},frame=single]
 float* problem_matrix =  (float*) rolex_malloc_robust ( N * sizeof (float), STRENGTH );

 void rolex_validate_robust ( void * problem_matrix);
\end{CodeExample}

\paragraph{Semantics} \hspace{0pt} \\
The \texttt{rolex\_malloc\_robust()} enables the programmer to request redundant copies of the memory block. The STRENGTH macro specifies the number of copies of the memory block. The pointer references to the replicated memory are also replicated at the source level, as well as any program statements that manipulate the memory. The \texttt{rolex\_validate\_robust()} routine initiates comparison and majority voting of the memory block. 

\subsubsection{Examples} 
Scientific applications employ data structures that heavily use pointer references and these are known to be highly sensitive to memory failures \cite{Aumann:1996:FCS}. Even single-bit upsets in pointer variables lead to invalid references, causing segmentation faults. Linear algebra methods, particularly those based on sparse problems, use structured formats such as dictionary of keys (DOK), list of lists (LIL), coordinate list (COO), compressed sparse row (CSR) or compressed sparse column (CSC) to refer to the non-zero elements (NNZ) of the sparse matrix. The bit precise correctness of such addressing structures and their computations is critical to application correctness. Using the {\tt robust} qualifiers and memory management routines for such variable state prevents potential error states arising due to bit corruptions since they are detected, or even corrected, before they lead to application failure due to invalid references. These robustness-based extensions may serve application-level variables that affect the program control flow, such as loop condition and if-else condition variables, which also demand bit-precise correctness. 

The robust directives may be applied to application phases whose reliability is critical to the application outcome. In molecular dynamics simulations, the correctness of the pairwise force calculation between the particles is critical for maintaining the numerical stability of the simulation. The directives may serve to provide in-situ detection and correction for these application phases, by leveraging the anti-symmetric property of the forces (for particles i and j, F$_{ij}$ = - F$_{ji}$ ) \cite{Yajnik:1994}. Linear solver methods, such as the FT-GMRES algorithm \cite{Hoemmen:2011} and the self-stabilizing conjugate gradient method \cite{Sao:2013}, permit partitioning of the algorithm into \textit{reliable} and \textit{unreliable} phases. In such a selective reliability model of execution, the correctness of the \textit{reliable} phases can be guaranteed through the redundancy-based error detection/correction semantics provided by the Rolex robust directives.

%% file: 04_c_Syntax_Semantics_AMELIORATION.tex
\subsection{Amelioration-based Extensions}
The \textbf{amelioration}-based language extensions are used to specify how data variables or code block executions may be repaired during program execution. The knowledge is based on algorithmic features of the application that allow the mitigation of the effects of errors on the program state. These methods compensate for the presence of errors by either maintaining encoding information on the variables, or by reconstructing incorrect values by interpolating from neighboring values. The amelioration approaches \cite{Hukerikar:HPEC:2015} may cause limited information loss, which may be acceptable to the user, but they seek to keep the application running towards solution rather than allow an error result in catastrophic failure of the application. 

\subsubsection{Type Qualifiers}
\paragraph{Syntax} \hspace{0pt} \\
The \texttt{heal} type qualifier enables amelioration through the association of a routine that may be invoked to repair anomalies in the annotated data structure. The \texttt{heal} may be applied to declarations of primitive as well as compound data structures. The syntax for the qualifier is:
\begin{CodeExample} [label = {lst:Example10},frame=single]
 heal (recovery_func()) float* matrix_A[N][N];
\end{CodeExample}

\paragraph{Semantics} \hspace{0pt} \\
The reference to the recovery function specified in the heal qualifier for the identifier in the type declaration is maintained by the runtime system. When the runtime receives an error notification for the heal qualified object, it invokes an event handler function with the recovery function pointer as argument. If the recovery function is able to repair the data structure, the runtime resumes the application process. The type qualifier provides error containment and recovery capabilities for the qualified object. 

\subsubsection{Directives}
\paragraph{Syntax} \hspace{0pt} \\
The amelioration-based directives provide limited localized recovery for regions of computation that are contained in the structured block following the directive and the associated data structures. The syntax for the amelioration directives is:
\begin{CodeExample}[label = {lst:Example11},frame=single]
#pragma rolex recover-rollback reinitialize ( variable_list )
{    /* code block *     }

#pragma rolex recover-rollforward reinitialize ( variable_list )
{    /* code block *     }

#pragma rolex recover-rollback ameliorate ( recovery_func() )
{     /* code block */   }

#pragma rolex recover-rollforward ameliorate ( recovery_func() )
{     /* code block */   }
\end{CodeExample}

These directives permit more flexible recovery of the variable state in addition to the roll-forward and roll-back capabilities. The list in the \texttt{reinitialize} and \texttt{ameliorate} clauses include variable identifiers, an expression list, or a user-defined recovery\_func(). 

\paragraph{Semantics} \hspace{0pt} \\
When the error notification to the runtime system finds that the error location is mapped to the program code contained in the structured block, or on the data variables manipulated by the statements in the block, the runtime initiates the recovery. This entails restoring the variable state for the variable identifiers specified in the \texttt{reinitialize} clause. When the recovery of variable state needs to be more nuanced, the runtime invokes a recovery function through an event handler. The runtime also affects a roll-back (re-entry of the code block) or a roll-forward (resume execution at the end of the code block). The amelioration directives support error containment as well as flexible recovery of the computation and variable state. 
 
\subsubsection{Runtime Library Routines}
\paragraph{Syntax} \hspace{0pt} \\
The library routines for memory allocation that support fault amelioration have the following APIs:
\begin{CodeExample}[label = {lst:Example12},frame=single]
float* problem_matrix =  (float*) rolex_malloc_repairable ( N * sizeof (float), checksum_func_pointer );

 void rolex_ameliorate_heal  ( void* problem_matrix );
\end{CodeExample}

The \texttt{rolex\_malloc\_repairable()} routine accepts a size argument and a pointer reference to a user-defined recovery function, which is registered with the runtime system when the memory block is allocated. The routine {\tt rolex\_ameliorate\_heal()} for the invocation of the recovery method only requires a reference to the memory block.

\paragraph{Semantics} \hspace{0pt} \\
When an error is detected on the memory block, the runtime invokes the recovery function through an event handler routine. When the recovery function is able to \textit{heal} the memory block, the runtime allows the application execution to resume. In case the recovery function is unable to correct the error, the runtime gracefully terminates the application process. The runtime library routine {\tt rolex\_ameliorate\_heal()} may be inserted in the application code to explicitly invoke the recovery function.

\subsubsection{Examples} 
Linear algebra methods that use dense matrix structures may maintain redundant information using checksum schemes to detect and correct perturbations. The checksum approach for amelioration is useful for a variety of matrix-based operations including matrix-matrix multiplication, Cholesky, LU and QR factorization methods. Sparse matrix-based problems, low overhead error detection and correction is possible by leveraging the structural properties of the matrix (diagonal, banded diagonal, block diagonal) using techniques such as approximate random (AR) checking and approximate clustered (AC) checking \cite{Sloan:2012}. These algorithm-based methods may be associated with the memory allocated for the matrix data structures using the Rolex amelioration type qualifier or memory allocation routine. 

Linear solvers based on iterative methods may be recovered from errors by replaying iterations. The amelioration directives support such recovery through roll-forward and roll-back semantics and clauses to re-initialize or repair the variable state, which enables any incorrect iterations to be discarded and keeps an iterative solver on the path to correct completion. Such partial recomputation techniques have been demonstrated to be viable error recovery methods for various linear algebra methods \cite{Sloan:2013}. Recovery may also be possible through lossy methods. For example, errors in the intermediate solution of Krylov subspace solvers may be recovered using interpolation of neighboring error-free values. The least-squares linear interpolation method has been demonstrated to be effective while maintaining the monotonic decrease in the residual norm \cite{Agullo:2013}. In the Hartree-Fock algorithm, heuristic knowledge is used to develop bounds for the data values. For the orthonormalization vector, density matrix, matrix exponential and orbital transformation structures, exact bounds conditions are known whereas data values for which sharp bounds are not known, such as the Fock matrix, a heuristic bound may be defined \cite{vanDam:2013}. Error states in data values are ameliorated by replacing them with reasonable values within these heuristic bounds. The Rolex amelioration constructs allow such knowledge to be conveniently embedded in the application code.

%% file: 05_Compiler_Runtime.tex
\section{Compiler and Runtime Support for Rolex}
\label{sec:Compiler_Runtime}

\subsection{Compiler Infrastructure}
The compiler infrastructure is a key intermediary that propagates the fault-resilience knowledge expressed by the programmer to the generated target code and runtime system. We have developed a compiler front-end, based on the ROSE compiler infrastructure \cite{ROSE:Compiler}, which parses the qualifiers and directives to generate code that is equipped with the resilience capabilities specified by the Rolex constructs. The front-end parses the resilience knowledge into a \textit{profile} file that is used by the runtime system. The front-end also performs source-to-source code transformations, which entails insertion of statements (using base language (C/C++) constructs) that permit the application to manage error states during execution in collaboration with Rolex runtime library (RTL) routines. A native C/C++ compiler may still be used to generate code for the target platform. The two-stage compilation process enables incorporating the resilience oriented transformations in the front-end while leveraging standard C/C++ compiler infrastructures to generate the target platform code. The modular approach permits selective compilation of resilience features through the use of compiler flags or, even bypassing the front-end compilation phase altogether. 

The front-end compiler parses all the Rolex qualified declarations in the program code in a single pass. For \texttt{tolerant} qualified objects, the compiler produces detection and correction masks based on the bit-level representation of the object type, which are included in the resilience profile file. For the \texttt{robust} qualified objects, the Rolex front-end duplicates/triplicates the declarations of the variables. It also traverses the uniform abstract syntax tree (AST) to discover the statements that perform operations on the \texttt{robust} qualified variables and inserts identical redundant statements for the replicated object copies and statements for comparison of the replicated variable values. For the \texttt{heal} type qualifier, a call to a RTL routine is added in order to register the recovery routine as a callback handler function. 

The front-end compiler pass also processes Rolex directives: it creates computational blocks for which the error detection, containment and correction behavior is explicitly defined. The front-end \textit{outlines} the statement list in the structured block that follows the Rolex directive into a new function. The original code block is replaced with a call to the outlined function. The front-end inserts calls to Rolex RTL routines, which affect roll-forward and roll-back semantics as well as support data scoping, preservation and restoration, prior to and after the call to the outlined function. The compiler also adds internal control variables (ICV), which are initialized and manipulated by the runtime to control the behavior of the outlined function. 

\subsection{Runtime Inference System}
In order to support a resilient execution environment, the runtime system manages the outcome of the error states in the application process. The runtime system maintains a resilience knowledge base, called the \textit{Dynamic Resilience Map (DRM)}, which contains the list of Rolex annotated data structures, their address offset in the address space and error-management strategies. The rules for error detection, containment and recovery strategies are those inferred from the Rolex annotations in the program source and parsed by the compiler into the profile file. These are populated into the DRM at the commencement of the application process execution. DRM entries are also dynamically added, removed and modified through the runtime library routines during the application execution. The runtime also provides an interface to the compiler front-end, which consists of RTL routines that are visible only to the compiler framework. The calls to these routines are associated with the outlined structured blocks. Table \ref{table:rolex-lib-routines} summarizes the Rolex RTL routines and their capabilities. 

\begin{table}
\begin{center}
 \begin{tabular}{r p{7cm}}
 \hline
 \textbf{Rolex library routine} & \textbf{Capability} \\ 
 \hline
 {\tt \_\_rolex\_initialize()}          & Initialization of runtime, allocation and population of the DRM\\
 {\tt \_\_rolex\_finalize()}            & Clean up of DRM and termination of runtime system\\
 {\tt \_\_rolex\_preserve\_state()}     & Preserve program's current state and environment\\ 
 {\tt \_\_rolex\_restore\_state()}      & Restore previously saved program state\\ 
 {\tt \_\_rolex\_jmp\_fwd()}            & Jump to pre-defined forward reference point and resume execution\\ 
 {\tt \_\_rolex\_jmp\_back()}           & Jump to pre-defined previous reference point and resume execution\\
 {\tt \_\_rolex\_create\_checkpoint()}  & Save state of variables listed in args\\
 {\tt \_\_rolex\_restore\_checkpoint()} & Restore state of variables from maintained copy in runtime\\
 {\tt \_\_rolex\_copy()}                & Duplicate the variable arg in the runtime\\
 {\tt \_\_rolex\_register()}            & Register program object in the DRM and default response\\
 {\tt \_\_rolex\_deregister()}          & Deregister program object from DRM\\
 {\tt \_\_rolex\_compare()}             & Compare memory of arg pointers\\
 \hline  
 \end{tabular}
 \caption{Rolex runtime library routines}\label{table:rolex-lib-routines}
\end{center}
\end{table}

\begin{figure} [tp]
\centering
\includegraphics[width=205mm, height=130mm, angle=90]{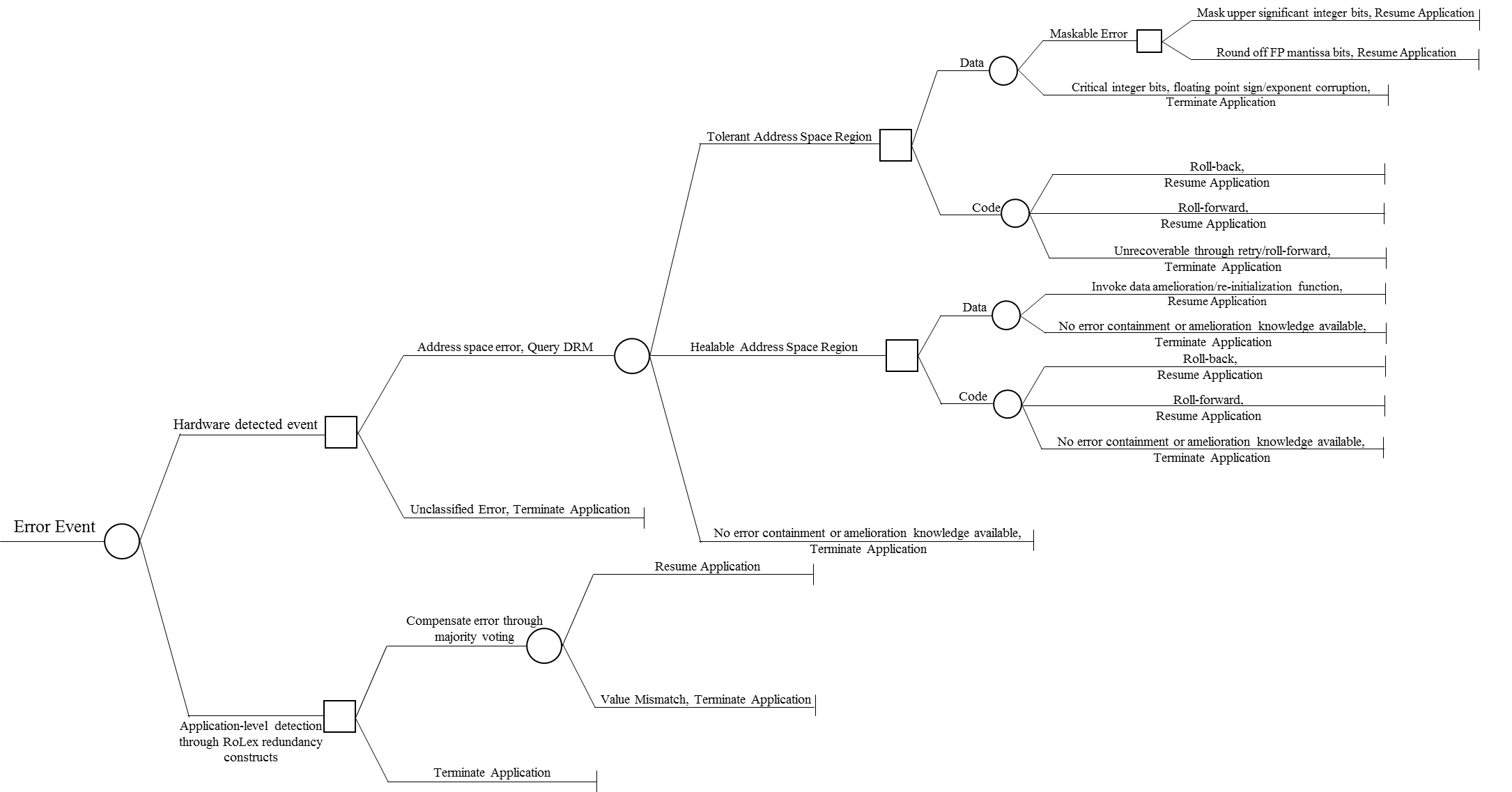}
\caption{Decision tree for error management by runtime inference system}
\label{Fig:Decision-Tree-DRM}
\end{figure}

When the runtime is notified of the presence of an error state in the application address space, it queries the DRM to find the specific application-level construct that is in error state. Based on the application construct in error state and the error management knowledge available in the DRM, the runtime invokes the appropriate RTL routines that seek to compensate for the perturbations in the variable state and rolls back or rolls forward the execution, if required. When the runtime is able to account for the error states, it allows the application process to resume execution in partially/fully restored computational state. The runtime actions are inferred by traversing the decision tree in Figure \ref{Fig:Decision-Tree-DRM}, which is constructed using the Rolex annotations on the various program constructs. The traversal provides the runtime with defintive rules to manage specific error states that may arise during the application program execution. When no error management knowledge is available for an application-level construct in the DRM, the runtime gracefully terminates the application process.

\subsection{Workflow of a Resilient Execution Environment}

\begin{figure} [tp]
\centering
\includegraphics[width=\linewidth, height=55mm]{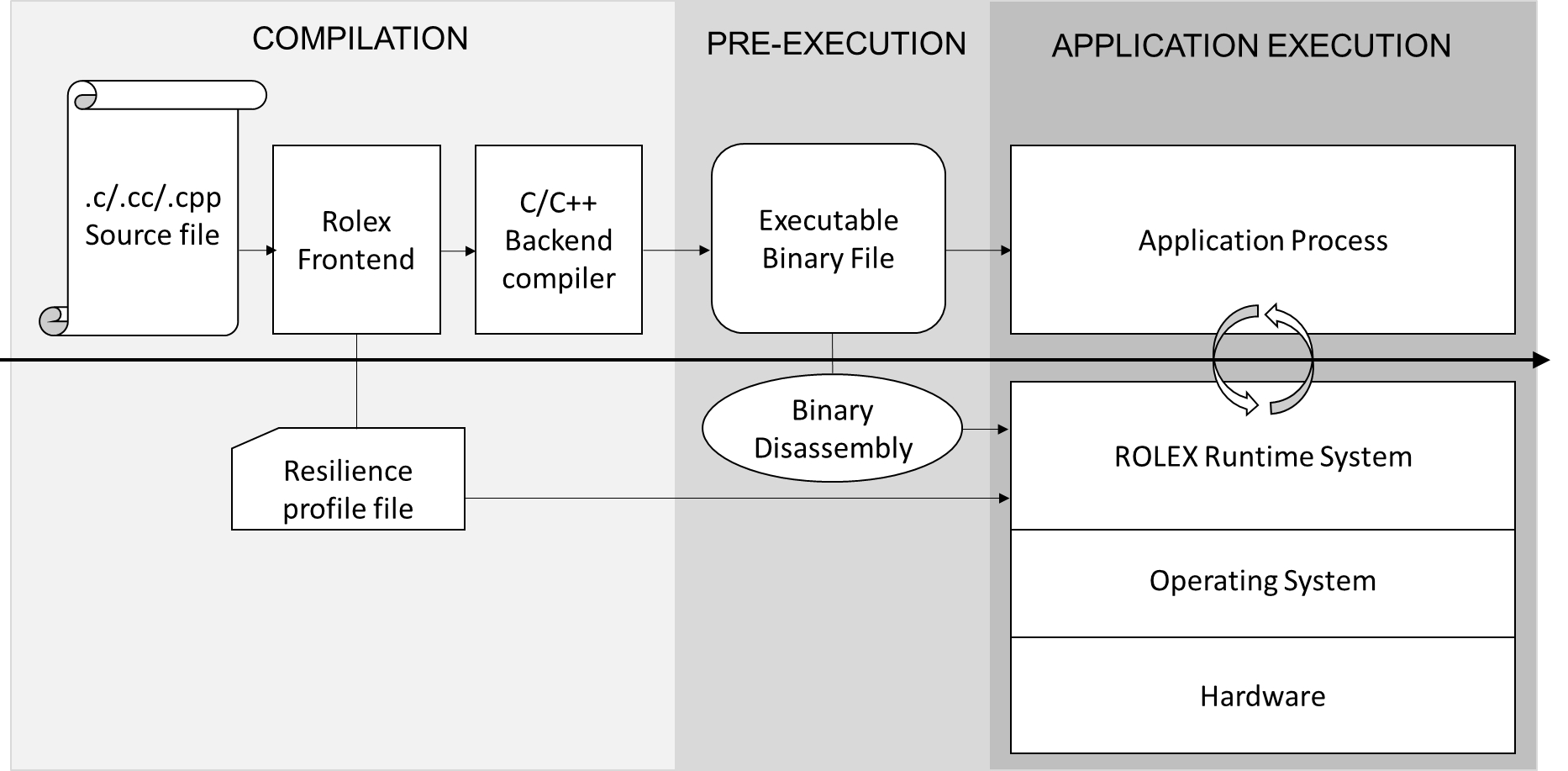}
\caption{Overview of application compilation and execution with Rolex}
\label{Fig:Sys-Overview}
\end{figure}

With the incorporation of Rolex, we allow several changes to the programming model and the execution environment, which are captured by Figure \ref{Fig:Sys-Overview}. When HPC application codes are annotated with the Rolex qualifiers and pragma directives, the compiler parses these extensions and introduces source-level transformations in the C/C++ application program code. The restructuring of the application source code to incorporate Rolex-driven resiliency features introduces additional declarations of redundant variables, outlining of blocks of code and creation of additional functions, and the installation of handler functions. Therefore, the program control flow and function call graph may be different from that intended by the application programmer, yet these modifications are transparent to the user. Additionally, when the application program is executed, we include a pre-execution stage where the linkages of the application-level constructs from the compiled binary are discovered through a binary disassembly library. During this phase the DRM is also populated with the address offsets and error-handling actions. 

In the current HPC execution models, the presence of a hardware detected error causes a machine check exception which raises an interrupt to the operating system. When the error state is uncorrectable, the kernel enters panic mode which leads to node shutdown. Therefore, all errors lead to failure and these are dealt with in failstop manner. With the support of the Rolex-based programming model, our execution environment includes a runtime inference system. The runtime is linked with the application code. The operating system contains a kernel module that intercepts the interrupts and passes them into the user space, i.e., to the runtime system through the signaling mechanism. The runtime contains a signal handler that contains the logic to query the DRM and to determine the best recourse for dealing with the error state. The runtime's RTL interface offers a well-defined API to augment the DRM knowledge base, which enables the runtime to affect error detection, containment and masking on application constructs. When the error state can be tolerated or ameliorated, the runtime allows the application execution to resume using the knowledge in the DRM. When no knowledge can be inferred, the runtime terminates the application, as is the norm for unrecoverable errors in current systems. Since the error-handling component of the runtime system is interrupt-driven, the runtime system does not add significant overhead to the application performance during error free execution.

The Rolex-based programming model makes the HPC applications fault-aware as well as fault-tolerant by imposing strict and relaxed reliability different on regions of the application state. Rolex enables an execution model in which there is an active interchange of error information between layers of the system stack. This prevents each error instance from causing a fatal application crash by reasoning about the significance of the error using the programmer's knowledge on the application's correctness expectations.

%% file: 06_Experimental_Eval.tex
\section{Experimental Evaluation}
\label{sec:Experimental_Eval}

\subsection{Accelerated Fault Injection Experiments}
In order to experimentally evaluate the benefits of using Rolex to describe the resilience properties of scientific application codes, we perform a set of accelerated fault injection tests. We use dynamic software-based fault injections into application processes and observe their impact on the application's outcome - whether Rolex enables the application to run to completion and whether the results produced are within reasonable bounds of a correct answer. 
For each application code, we use five fault injection rates: 1 fault/15 minutes, 1 fault/10 minutes, 1 fault/5 minutes, 1 fault/2 minutes and 1 fault/1 minute. By adjusting the input problem sizes, the execution time of each application run is adjusted to be greater than 20 minutes; this ensures that the application process execution experiences at most 1, 2, 4, 10 and 20 faults per run. With the fault rates that we have selected, the effective mean-time-to-error of the application process is set to 15, 10, 5, 2, and 1 minute(s). In comparison to the fault rates observed on production HPC systems today these error rates are extremely high. These rates are also significantly higher than most reasonable projections for exascale-class systems based on technology roadmaps. However, these experimental fault rates were chosen to validate the dependability of the application processes and the efficacy of a Rolex-based programming environment. They also provide insights into the precise behavior of the application in the presence of faults. Also, several error modes that are unseen today might emerge in future systems and these accelerated tests serve as stress tests for such scenarios.

Since some of the extensions only support tolerance and amelioration semantics, they rely on hardware-based detection mechanisms. Other Rolex features provide implicit error detection. Therefore, the type of the fault injected, i.e., whether it results in a detected memory error or a silent data corruption, depends on the type of Rolex extension being evaluated. We have developed a flexible software-based fault injection framework that simulates the different error behaviors. The fault injection framework is non-intrusive, i.e., it runs independently from the application process and does not require  modification of the application program code, or compiler-based insertion of additional instructions. It simulates a hardware interrupt by passing a signal to the application process. The fault injection framework maintains a mapping of the address space of the application process and the offsets for the various application-level constructs and can inject faults into any region of the active address space. The fault site selection may be random or may target specific application constructs. The faults injection entails flipping the bits at the selected fault site in the application address space.    

We evaluate the application resilience of the scientific codes by opportunistically annotating their source with the Rolex-type qualifiers, directives and runtime library routines to suit the inherent resilience properties of the code. The code is compiled with our ROSE-based front-end compiler and then with the GCC compiler infrastructure and is linked with the Rolex runtime library. The application binaries are executed in a Linux-based cluster environment. For each fault injection rate, the application run is performed 10,000 times each with randomly selected fault injection sites. 

\subsubsection{Enabling Tolerance Using Rolex Extensions}
\begin{figure*} [t]
\centering
\includegraphics[width=\linewidth,height=50mm]{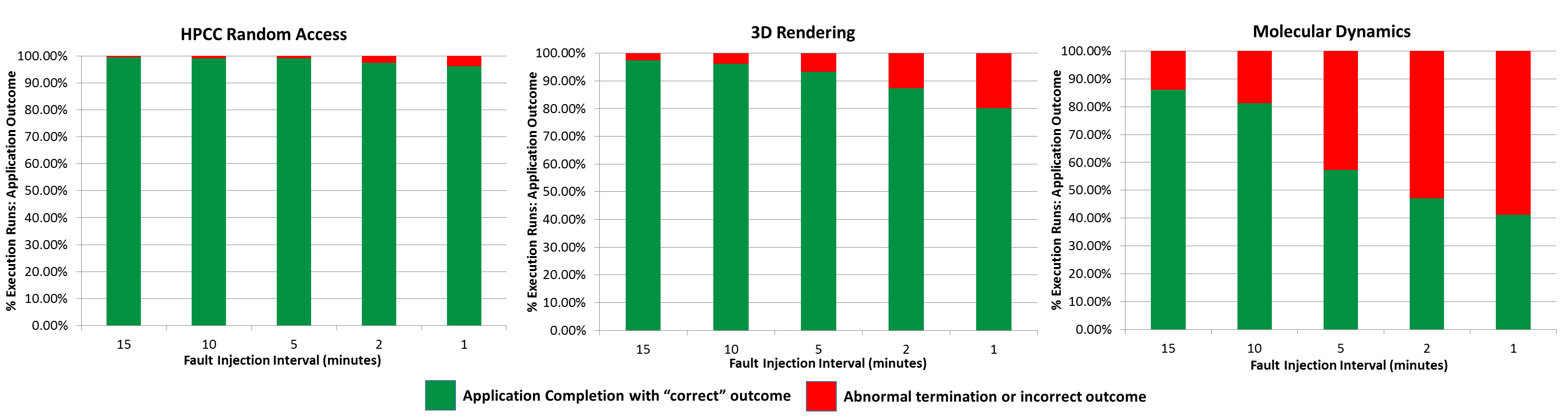}
\caption{Evaluation of tolerance Rolex extensions: Accelerated fault injection results}
\label{Fig:Results-Tolerance}
\end{figure*}

Rolex extensions for error tolerance support elision semantics or, provide value coercion, but seek to keep the application process running towards completion. Since these extensions depend on hardware-based detection mechanisms, the injected faults simulate system memory errors that manifest themselves as ECC SECDED errors (detected but unrecoverable by hardware-based ECC), whose notification is passed into the runtime system. Based on the location of the error, there are only two possible outcomes: compensation for the presence of the error (through error elision, masking the affected bits of the variables, or roll-forward/roll-back of the execution), or termination of the application to prevent further corruption. We simulate SECDED errors by raising a signal when the fault injection framework perturbs bits in the address space. We demonstrate error tolerance through Rolex for the following three codes:
\begin{itemize}

\item \textbf{HPCC Random Access:}
The benchmark was originally designed to model a vectorized application and allows the same address to appear twice in a gather/scatter operation and therefore fails to guarantee sequential consistency. Due to this property, the benchmark is explicitly tolerant to the presence of errors in its HPCC Table array. The computational kernel performs repeated pseudorandom updates. We allocate the HPCC Table array structure using the {\tt rolex\_malloc\_tolerant()} runtime library routine to support error elision semantics on the memory region corresponding to the HPCC Table.

\item \textbf{3D Rendering Application:} The application converts a 3D model of a scene into a 2D screen representation. The final rendered scene is written to a frame buffer which is declared as a 2-D array in our test code. In order to ignore the presence of perturbations in the frame buffer, we qualify its declaration with the {\tt tolerant} type qualifier. For these application runs, the measure of correct completion is an execution that completes and renders the scene in which fewer than 5\% of pixel values are perturbed beyond a local characteristic threshold value. 

\item \textbf{Molecular Dynamics Simulation:}
This simulation is based on time-stepping algorithm and contains floating-point array structures for the particle position, velocity and acceleration. These are calculated every time interval and it has been demonstrated that these coercing these vectors into lower precision does not affect the stability of the simulation over a large number of time steps. We qualify their declaration with the tolerant type qualifier and use the PRECISION construct to apply relaxed precision for the lower 26 bits of the mantissa (when declared using double-precision type). We monitor the properties of the complete system, including total energy and pressure, to determine the validity of a simulation run.

\end{itemize}

Figure \ref{Fig:Results-Tolerance} summarizes the results of these fault injection experiments. These results show the percentage of the total application runs that complete correctly despite the injected errors versus those that end fatally. In the Random Access benchmark, the memory footprint of the computational kernel that performs the pseudorandom updates is extremely small in comparison to the HPCC\_Table array, which occupies 50\% of the system memory and allocated with the tolerant version of the malloc routine. Therefore, upto 99\% of the execution runs converge - even for an error rate as high as 1 fault per minute. Similar resiliency features are demonstrated by the 3D rendering application in which the dominant portion of the active memory footprint is the integer type frame buffer array, which is declared with the {\tt tolerant} qualifier. 
For the molecular dynamics simulations, the only resilience property exposed through Rolex is the relaxed precision on the position, velocity and acceleration arrays. This supports error tolerance on only a limited fraction of the total active address space. Accordingly as many as 85\% of the application runs converge correctly for a fault rate of 1 fault/5 minutes; the survival rate drops rapidly in the presence of higher fault rates. The ``failed" simulations include runs that terminate abnormally as well as completed runs in which with total energy and/or pressure of the system diverges outside $\pm$5\% of a fault-free run.   

\subsubsection{Enabling Robustness Using Rolex Extensions}
\begin{figure*} [t]
\centering
\includegraphics[width=\linewidth,height=60mm]{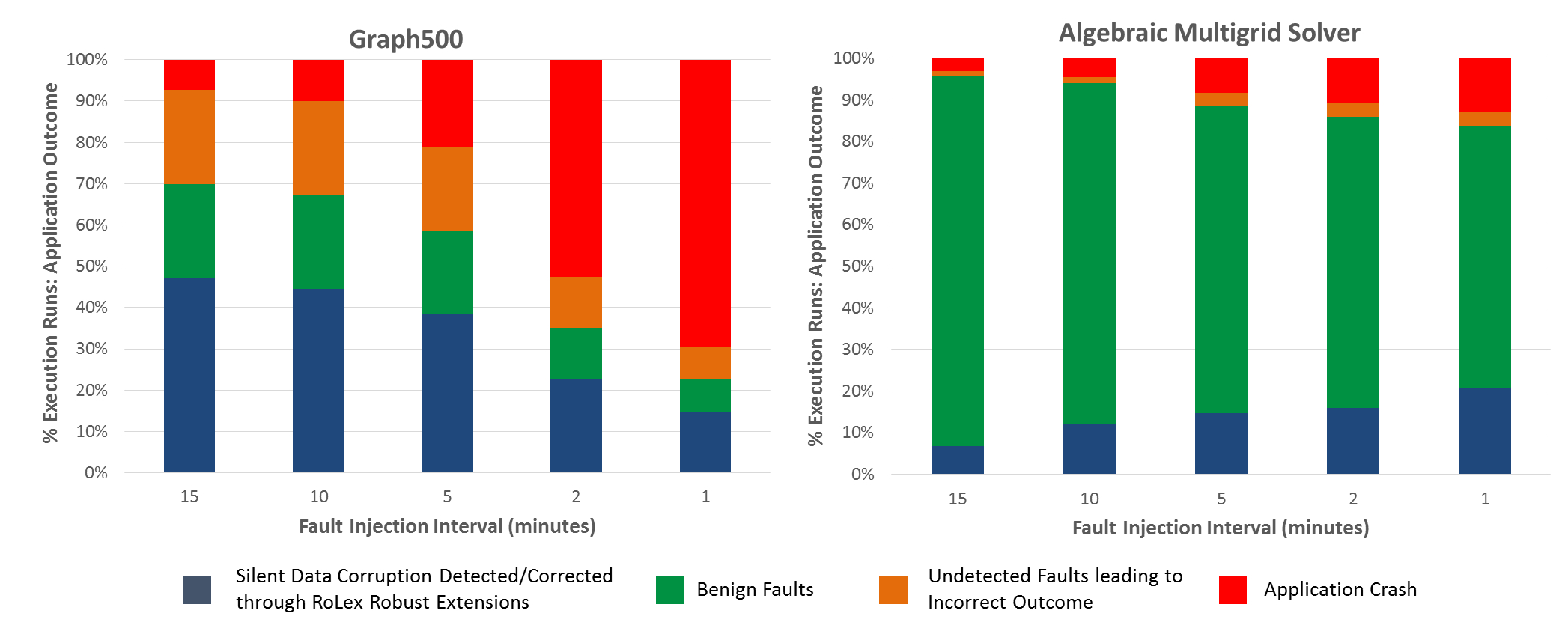}
\caption{Evaluation of robustness Rolex extensions: Accelerated fault injection results}
\label{Fig:Results-Robustness}
\end{figure*}

The Rolex extensions for robustness provide error detection and correction semantics through the use of redundancy. Since application-level error detection is often implicitly supported for such robust annotated application constructs, we make no assumptions about hardware-level detection and notification mechanisms. For these experiments, the fault injection framework simulates silent data corruptions (SDC). For these injections, the target application process is intercepted and bit-flip perturbations are introduced at the fault site. No notification is raised to the runtime system, and the fault injection framework allows the application process to resume execution. We consider four possible outcomes of a bit corruption injected in the application address space:
\begin{itemize}
\item \textbf{Silent data corruptions} that are detected using the redundancy injected into the application code.
\item \textbf{Benign faults} that remain in the program state until the conclusion of the execution, but do not affect the correctness of the outcome.
\item \textbf{Undetected faults} in the application state cause errors but these fall outside the coverage provided by the Rolex constructs.
\item \textbf{Application crash} that occurs when the injected perturbation affects part of program state mapped to the computational environment.
\end{itemize}
For the fault injection runs, we observe the propagation of the fault after injection until the application completes, or terminates. We apply the robustness extensions on the following two codes using Rolex:
\begin{itemize}

\item \textbf{Graph500 Breadth-First-Search:} This unstructured, integer-oriented benchmark is based on the graph abstraction and the code contains several pointer references that represent the graph edges and vertices. The correctness of these pointers is critical to the successful completion of an application run since any perturbations on these lead to usually lead to illegal address accesses and a fatal crash of the application process. We qualify all the pointer declarations for the graph edges and vertices with the robust qualifier in the Graph500 Breadth-First-Search (Kernel 2) code \cite{Hukerikar:HPEC:2013}. 

\item \textbf{Algebraic Multigrid Solver:} Each multigrid iteration of the linear solver, referred to as a ``V-cycle," consists of smoothing, restriction and interpolation stages during which the algorithm starts with a fine grid, restricts to a coarser grid and then interpolates to a fine grid again. The intermediate solution grids are known to tolerate errors at the cost of needing additional V-cycles to converge to the correct solution. However, the algorithm is also sensitive to pointer variable corruptions. We apply the robust qualifier for each pointer variable declaration in code. Additionally, we allocate the intermediate solution grids using the \texttt{rolex\_malloc\_tolerant()} routine.

\end{itemize}

The Figure \ref{Fig:Results-Robustness} illustrates the distribution of the application outcomes for each fault injected. The Graph500 BFS algorithm contains a large number of pointer-related computations to traverse the graph edges. It is possible to detect and correct the corruptions in the pointer arithmetic for almost 50\% of all corruptions injected for a fault interval of 15 minutes. Since the number of visits for each vertex is fixed in the BFS algorithm, the memory for these vertices and their pointers are not used as the application progresses. Silent corruptions on these regions of the application address space are benign. Other parts of the computational environment as well as the graph vertex data elements contain no error management knowledge. When the injected faults hit these regions the application fails. Therefore, a majority of injected faults are fatal to the application at fault intervals of 1 and 2 since the Rolex fault coverage only protects the pointer variable state. The AMG code demonstrates a different resilience behavior since the address space dedicated to the inherently resilient intermediate solution grids is a significant part of the total address space. Therefore, although the Rolex constructs only provide coverage for the pointer variables, a majority of the injected silent faults still turn out benign since the resulting error in the intermediate state is refined by the iterative nature of the algorithm.

\subsubsection{Enabling Amelioration Using Rolex Extensions}
\begin{figure*} [t]
\centering
\includegraphics[width=\linewidth,height=50mm]{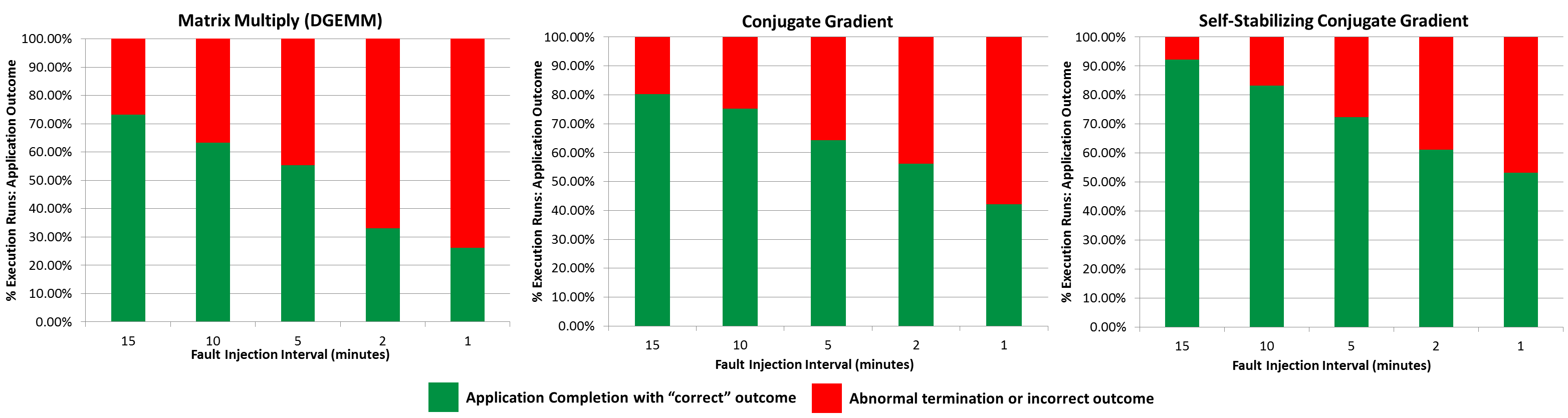}
\caption{Evaluation of amelioration Rolex extensions: Accelerated fault injection results}
\label{Fig:Results-Amelioration}
\end{figure*}
The Rolex extensions for amelioration enable recovery of the application's variable or computational state by using well-known algorithm-based fault tolerance methods. The extensions must be supported by hardware-based detection mechanisms. Since these extensions associate a recovery function with a data structure or computation, there are only two possible outcomes for each fault detected: the application state is repaired by the recovery function, which permits the application execution to resume or, the application must terminate since the recovery function is insufficient to repair the corruption. We demonstrate fault amelioration using Rolex constructs for the following codes:    
\begin{itemize}

\item \textbf{Matrix-Matrix Multiplication:}
In the DGEMM code, calculating row and column-wise checksums is a well-known solution to detect and correct corruptions in the matrix. We define functions that maintain the row and column checksums for the operand matrices whose reference is passed to the {\tt rolex\_malloc\_repairable()} library routine. If the matrix declaration is static the recovery function may be included in a {\tt heal} type qualifier.

\item \textbf{Conjugate Gradient Solver:}
For the CG solver, the matrix is allocated by the library routine \texttt{rolex\allowbreak\_malloc\allowbreak\_repairable()}. The pointer to a function that maintains checksums of the matrix is passed to this routine. Additionally, we leverage the iterative property of the CG algorithm by including the  CG iteration step in the {\tt \#pragma rolex roll-forward} amelioration directive and associate the checksum routine with the directive. This allows faulty iterations to be discarded and validating the correctness of the operand matrix upon roll-forward.

\item \textbf{Self-Stabilizing Conjugate Gradient:}
The self-stabilizing version of CG offers a correction step that restores the stability of the algorithm when it is affected by errors. This correction step is included in a recovery function whose reference is included in the {\tt ameliorate} clause of a directive. The CG iteration steps are included in the amelioration directive {\tt \#pragma rolex roll-back}. The roll-back capability allows the most recent faulty CG iteration to be discarded and the recovery to be invoked. 
\end{itemize}

Figure \ref{Fig:Results-Amelioration} summarizes the results of these experiments. For the DGEMM code, the checksum-based amelioration is applicable for only the static state in the application address space, i.e., the operand matrices that are initialized at the beginning and whose values do not change throughout the execution. We have not applied any Rolex construct on the dynamic state, i.e., the result matrix. With this fault coverage, 75\% of all executions converge correctly for the fault rate that injects an error every 15 minutes, but only 27\% complete correctly at the accelerated rate of 1 error per minute in which case as many as 20 unrecoverable errors are injected into the process state. The inclusion of Rolex constructs to the CG solver yields a better resilience characteristic than DGEMM for similar fault intervals. This is because in addition to the checksum-based error detection/correction on the operand matrices, the iterative nature of the algorithm permits incorrect computation to be recovered. Due to the enhanced address space fault coverage through Rolex in CG codes the application demonstrates a better completion rate than DGEMM, even at higher fault rates. The SS-CG contains a correction step that is designed to restore the stability of the algorithm. This permits relaxation of the reliability requirements for the CG iterations. Therefore, a larger percentage of executions of the SS-CG converge correctly, in comparison to CG, for similar fault rates.

\subsection{Performance Evaluation}
\begin{figure*} [t]
\centering
\includegraphics[width=\linewidth,height=80mm]{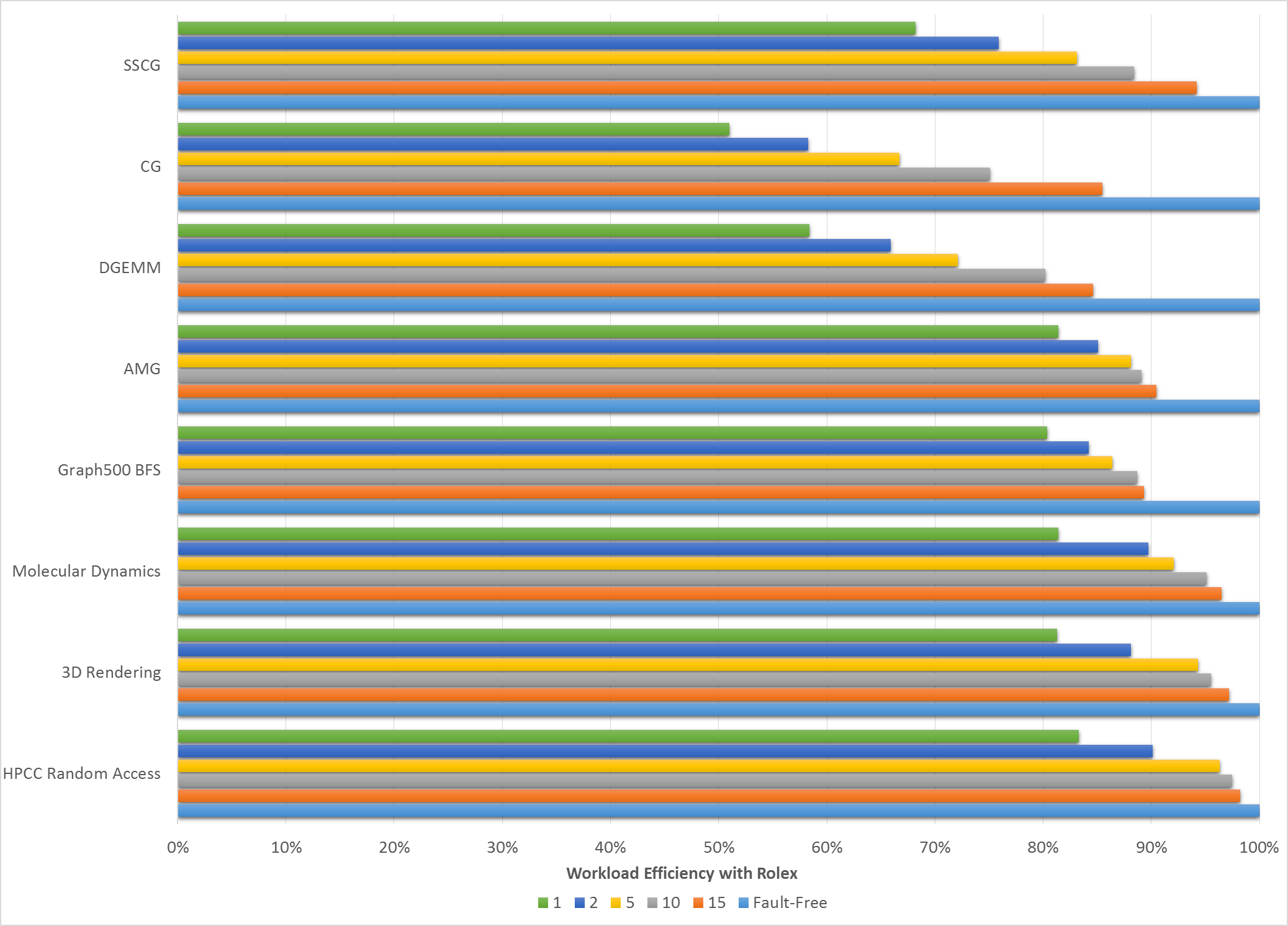}
\caption{Performance Evaluation of Rolex: Workload Efficiencies}
\label{Fig:Results-Performance}
\end{figure*}

We evaluate the overhead of embedding the resilience knowledge using Rolex for each of the application codes. With the introduction of Rolex constructs in the applications' source code, the overhead is introduced by compiler-inserted statements as well as runtime library routines. Additionally, the response to each type of error depends on its context, i.e., its location in the address space and the knowledge available in the runtime's DRM. Therefore, we evaluate the performance impact by comparing the workload efficiency which is the ratio of the ideal time-to-solution on a fault-free execution run to the actual running time in the presence of faults:
\begin{equation}
Efficiency = \frac{t_{fault-free}}{t_{actual-in-presence-faults}}
\end{equation}
The difference between t$_{fault-free}$ and t$_{actual-with-faults}$ is the overhead associated with dealing with faults by the Rolex runtime system. This includes the time for fault detection, diagnosis and applying any recovery and compensation. We compile each application code to two different binary versions: a binary with Rolex, compiled using our front-end source-to-source compiler followed by a regular GCC compiler; and a version using only a GCC compiler. The binary version without Rolex is executed in a fault-free environment to measure the baseline execution time. The version containing Rolex is subjected to fault injection for which we measure the application's time to solution for runs that survive all the faults and reach correct completion. The execution times for each fault are averaged for the fraction of the 10,000 application runs that complete correctly. This allows examination of the overhead incurred by the compiler-based transformations as well as the overhead incurred by the runtime inference system. 

The results for the workload efficiency are summarized in Figure \ref{Fig:Results-Performance}. The overhead to manage errors in HPCC Random Access and the 3D rendering application are low because the runtime tolerates errors through elision and the size of the DRM is very small. Therefore, even for extremely high fault rates, the overhead is about 15\%. For the molecular dynamics simulation, the error tolerance is supported through value coercion on the position, velocity and acceleration vectors and this operation incurs a higher overhead than error elision. Consequently, the overhead for the largest fault interval is 4\% and as much as 19\% for the smallest fault interval.
The robustness-based constructs introduce redundancy through compiler-based transformations into the application source. However, since we only annotate the pointer variables in both the Graph500 BFS and AMG codes, there is a fixed overhead cost of about 10\% attributed to the redundant statements. The lower efficiency at higher fault rates may be attributed to the overhead in notifying the runtime system. The amelioration-based Rolex constructs demonstrate a significantly higher overhead compared to the tolerance and robustness extensions. However, much of this overhead may be attributed to the algorithmic amelioration functions rather than compiler and runtime overheads. The SS-CG offers the best efficiency among the codes that use the amelioration constructs since it requires only a stabilization step. The checksum operations are computationally expensive operations and therefore the efficiency of the DGEMM and CG codes are lower, particularly at higher fault rates when the checksum functions are invoked frequently.

%% file: 07_Related_Work.tex
\section{Related Work}
\label{sec:Related_Work}

HPC programmers have historically borne the burden of exploiting novel features in system architectures and execution models in the pursuit of performance. They usually rely on various extensions to high-level programming languages with the support of compilation techniques and runtime libraries. For example the OpenMP \cite{OpenMP:Spec} standard emerged in order to support shared memory multiprocessing programming in C, C++, and Fortran through a set of directives, library routines and environment variables. Similarly Berkeley's UPC effort \cite{Carlson:1999:UPC} also extends the C language with constructs that present the programmer with a single global partitioned global address space as the program runs on shared or distributed memory parallel systems. The Co-array Fortran (CAF) \cite{Numrich:1998:CFP} began as an extension of Fortran 95/2003 (and became part of the Fortran 2008 standard) to support the PGAS model for Fortran programs. NVIDIA's CUDA was derived from Brook \cite{Buck:2004:SIGGRAPH} which extended the C language with data-parallelism-oriented constructs that enabled the use of the graphics processing units (GPU) as streaming co-processors.

The support for fault tolerance capabilities through programming models-based approaches has been recently proposed and evaluated. Programming constructs called containment domains \cite{Chung:2011:SC} provide the application programmer with mechanisms to delineate computation that have transactional semantics. Upon execution of the code block, the results of the computation are checked for correctness and if the block's execution condition is not met, the results are discarded and the block may be re-executed. Similarly, language-level support for idempotent regions \cite{deKruijf:2012:PLDI} enables application programmers to specify "relax" blocks in C/C++ programs, which may be freely re-executed without checkpointed state or side-effects. The FaultTM scheme \cite{Yalcin:2010:PESPMA} requires an application programmer to define vulnerable sections of code which are executed by duplicate thread contexts. The original and the backup thread are executed as an atomic transaction, and their respective result values are compared before committing the result.
The Global View Resilience (GVR) project \cite{Fujita:2013:ASPLOS} provides annotations to create multiple snapshot versions of the application data, which enables recovery from failures by restoring the application state to a previous snapshot version.

%% file: 08_Conclusion.tex
\section{Conclusion}
\label{sec:Conclusion}

This paper presented a set of Resiliency-Oriented Language Extensions (Rolex) for expressing the error resilience properties of scientific HPC application codes at the language level. They are developed as extensions to existing programming languages such that they may succinctly capture a programmer's knowledge on the fault tolerance features of the application through type qualifiers, directives and library routines. The semantics of the language extensions enable application-level error detection, containment and masking. We have presented concrete examples of widely used scientific computational kernels in which encoding the resilience knowledge using Rolex enhances the application's error resilience. We described the compiler transformations that leverage the language extensions to incorporate further error resilience features in the application codes. These transformations are enabled by a front-end source-to-source compiler infrastructure. We described the compiler-runtime interface and the design and implementation of the runtime inference system.
We demonstrated that the combination of the language-level programming model extensions, which are tightly integrated with the compiler infrastructure and runtime system, provides an execution environment that facilitates cross-layer efforts for error detection, masking and recovery.
For HPC applications, these capabilities in turn lead to a substantial increase in the checkpointing interval and a reduction in redundant computation, both of which enable a reduction in the time and energy required to reliably solve the most demanding computational challenges.

%% file: A_Appendix_Grammar.tex
\section{Rolex Grammar}

This appendix shows the extensions to the base language grammar for C and C++ in order to support Rolex. 

\subsection{Rules for Resilience Type Qualifiers}

\begin{Code} [caption= {Rules for resilience type qualifiers}, label = {lst:TypeQualifierRules}]

declaration_specifiers : storage_class_specifier
                       | storage_class_specifier declaration_specifiers
                       | type_specifier
                       | type_specifier declaration_specifiers
                       | type_qualifier
                       | type_qualifier declaration_specifiers
                      ';'

storage_class_specifier : TYPEDEF | EXTERN | STATIC | AUTO | REGISTER ';'

type_specifier : VOID | CHAR | SHORT | INT | LONG | FLOAT | DOUBLE | SIGNED | UNSIGNED
               | struct/union_specifier | enum_specifier | TYPE_NAME
              ';'

type_qualifier : CONST
               | VOLATILE
               | resilience_type_qualifier
              ';'

resilience_type_qualifier : TOLERANT
              | TOLERANT '('  tolerance_limit  ')'
              | ROBUST '('  robust_strength  ')'
              | HEAL '(' function_declaration ')'
             ';'

tolerance_limit : PRECISION '=' CONSTANT
                | MAXIMUS   '=' CONSTANT

robust_strength: DETECT | CORRECT

\end{Code}

\subsection{Rules for Resilience Directives}
\begin{Code}[caption = {Rules for resilience directives},label = {lst:DirectiveRules}]

statement-list: statement
      | resilience-directive
      | statement-list statement
      | statement-list resilience-directive

statement : labeled_statement
          | compound_statement
          | expression_statement
          | selection_statement
          | iteration_statement
          | jump_statement
          | resilience-construct
	  | declaration-definition
	  | function-statement
          ';'

resilience-construct: rolex-redundancy-construct
                    | rolex-recovery-construct
		    | rolex-declare-construct

rolex-redundancy-construct: redundancy-directive structured-block

rolex-recovery-construct: recovery-directive structured-block

rolex-declare-construct: declare-directive function-statement

structured-block: statement

recovery-directive:#pragma rolex recover-rollback recovery-data-clause(opt) new-line
                   #pragma rolex recover-rollforward recovery-data-clause(opt) new-line

redundancy-directive: #pragma rolex robust robust-strength-clause redundancy-data-clause(opt) new-line

declare-directive: #pragma rolex declare resilient declare-resilience-clause failsafe-data-clause(opt) new-line

robust-strength-clause: DETECT | CORRECT

recovery-data-clause: data-default-clause
                    | data-private-clause
                    | data-share-clause
                    | data-reinitialize-clause
                    | data-ameliorate-clause

redundancy-data-clause:  data-default-clause
                      | data-private-clause
                      | data-share-clause
                      | data-compare-clause

failsafe-data-clause: fallback '(' variable-list ')'

data-default-clause: default '(' shared ')'
                   | default '(' none ')'

data-private-clause: private '(' variable-list ')'

data-share-clause: share '(' variable-list ')'

data-reinitialize-clause: reinitialize '(' variable-list ')'

data-ameliorate-clause: ameliorate '(' function_declaration ')'

data-compare-clause: compare '(' variable-list  ')'

declare-resilient-clause: retry
			| ignore 
                        | robust '('  robust_strength  ')'
\end{Code}

The redundancy directives enable error detection and/or correction for the computation contained in a structured block. The strength clause indicates whether dual or triple modular redundant execution must be applied. The recovery directives offer error containment since any fault that is activated leading to error state during the execution of the structured block is not allowed to propagate outside the block. Error recovery is performed by rolling forward or rolling back execution of the structured block.
The roll-forward and roll-back semantics on the structured code blocks require explicit specification of the data scoping to comply with the C/C++ memory consistency model. The rules for the data management and scoping clauses are also shown in Listing \ref{lst:DirectiveRules}. The clauses permit the variable state to be restored when execution is rolled forward or back. For the redundancy directives, the data clauses ensure that there are no races on the shared data. The declarative clauses in Rolex enable the creation of multiple versions of the associated function in order to support retry, ignore or redundant execution for the statements in the function body.